\documentclass[twocolumn]{aastex631}

\usepackage{tablefootnote}

\begin{document}

\title{X-ray Constraints on Wandering Black Holes in Stripped Galaxy Nuclei in the Halo of NGC 5128}

\author[0009-0006-4435-0418]{Samuel L. Feyan}
\affiliation{Center for Data Intensive and Time Domain Astronomy, Department of Physics and Astronomy,\\ Michigan State University, East Lansing, MI 48824, USA}
\affiliation{Department of Physics and Astronomy, Clemson University, Clemson, SC 29634, USA}
\author[0000-0003-1814-8620]{Ryan Urquhart}
\affiliation{Center for Data Intensive and Time Domain Astronomy, Department of Physics and Astronomy,\\ Michigan State University, East Lansing, MI 48824, USA}
\author[0000-0002-1468-9668]{Jay Strader}
\affiliation{Center for Data Intensive and Time Domain Astronomy, Department of Physics and Astronomy,\\ Michigan State University, East Lansing, MI 48824, USA}
\author[0000-0003-0248-5470]{Anil C. Seth}
\affiliation{Department of Physics and Astronomy, University of Utah\\
115 South 1400 East, Salt Lake City, UT 84112, USA}
\author[0000-0003-4102-380X]{David J. Sand}
\affil{Steward Observatory, University of Arizona, 933 North Cherry Avenue, Tucson, AZ 85721, USA}
\author[0000-0003-2352-3202]{Nelson Caldwell}
\affiliation{Center for Astrophysics, Harvard \& Smithsonian, 60 Garden Street, Cambridge, MA 02138, USA}
\author[0000-0002-1763-4128]{Denija Crnojevi\'{c}}
\affiliation{Department of Physics \& Astronomy, University of Tampa, 401 West Kennedy Boulevard, Tampa, FL 33606, USA}
\author[0000-0003-0234-3376]{Antoine Dumont}
\affil{Max Planck Institute for Astronomy, K{\"o}nigstuhl 17, 69117 Heidelberg, Germany}
\author[0000-0001-6215-0950]{Karina Voggel}
\affiliation{Universite de Strasbourg, CNRS, Observatoire astronomique de Strasbourg, UMR 7550, F-67000 Strasbourg, France}

\begin{abstract}

A subset of galaxies have dense nuclei, and when these galaxies are accreted and tidally stripped, the nuclei can masquerade as globular clusters in the halos of large galaxies. If these nuclei contain massive central black holes, some may accrete gas and become observable as active galactic nuclei. Previous studies have found that candidate stripped nuclei rarely host luminous X-ray sources, but these studies were typically restricted to both the most massive candidate nuclei and the most luminous X-ray sources. Here we use new and archival Chandra and XMM-Newton data to search for X-ray emission in a near-complete sample of massive globular clusters and candidate stripped nuclei in the nearest accessible elliptical galaxy, NGC 5128. This sample has the unique advantage that the candidate stripped nuclei are identified dynamically via elevated mass-to-light ratios. Our central result is that 5/22 ($23^{+11}_{-6}$\%) of the candidate stripped nuclei have X-ray sources down to a typical limit of $L_X \sim 5 \times 10^{36}$ erg s$^{-1}$, a fraction lower than or comparable to that among massive clusters with normal mass-to-light ratios (16/41; $39^{+8}_{-7}$\%). Hence we confirm and extend the result that nearly all X-ray sources in stripped nuclei are likely to be X-ray binaries rather than active galactic nuclei. If the candidate stripped nuclei have black holes of typical masses $\sim 2 \times 10^{5} M_{\odot}$ needed to explain their elevated mass-to-light ratios, then they have typical Eddington ratios of $\lesssim 2 \times 10^{-6}$. This suggests that it will be challenging to conduct an accretion census of wandering black holes around even nearby galaxies.

\end{abstract}

\section{Introduction}
\label{sec:intro}

A central prediction of hierarchical galaxy formation is the accretion of less massive galaxies by a more massive parent (\citealt{Somerville2015}). If the orbit of the accreted galaxy takes it close enough to the center of the parent, it can be tidally stripped, as observed in the Galaxy for dwarfs such as Sagittarius \citep{Ibata1994}. Since most galaxies with stellar masses from $10^{8}$ to $10^{10} M_{\odot}$ have dense nuclear star clusters with typical masses $10^{6}$--$10^{8} M_{\odot}$ \citep{Neumayer2020}, a straightforward expectation is that massive galaxies should be surrounded by a population of a few to many dense stripped galaxy nuclei \citep{Pfeffer2014,Tremmel2018}. 

The discovery of candidate stripped nuclei---historically called ``ultra-compact dwarfs" (UCDs)---bloomed as a subfield with the advent of large spectroscopic surveys of massive elliptical galaxies \citep{Hilker1999,Drinkwater2000,Phillipps2001,Drinkwater2003}, eventually augmented by size-based selection primarily using Hubble Space Telescope imaging (e.g., \citealt{Hasegan2005,Evstigneeva2007,2011AJ....142..199B}). These surveys identified UCDs as sources that were more massive and/or larger than typical globular clusters.

Spectacular confirmation of the stripping scenario came via the dynamical detection of a $2 \times 10^{7} M_{\odot}$ supermassive black hole in M60-UCD1 \citep{Seth2014}, one of the densest and most massive ($\gtrsim 10^{8} M_{\odot}$) UCDs known, consistent with being the tidally stripped remnant of a Milky Way-mass galaxy \citep{Strader2013}. A few more UCDs have been subsequently confirmed to also host supermassive black holes \citep{Ahn2017,Ahn2018,Afanasiev2018}, all in UCDs with stellar masses $\gtrsim 3 \times 10^{7} M_{\odot}$, far beyond the mass regime of normal globular clusters.

The census of the less massive UCDs, with stellar masses $\lesssim 10^7 M_{\odot}$, bears on multiple open questions in the assembly of galaxies and supermassive black holes. The stripped nuclei preserve a fossil record of tidally disrupted galaxies, with kinematic and chemical information that is challenging to obtain for more distant galaxies. They are also a promising route to constrain the occupation fraction of supermassive black holes for lower-mass galaxies, which informs seeding scenarios for black hole growth \citep{Volonteri2010}. If the occupation fraction is high in lower-mass galaxies, then the supermassive black hole number density in the local universe could be dominated by stripped nuclei \citep{Voggel2019}, and extreme or intermediate mass-ratio gravitational wave sources detectable by future gravitational wave observatories would be more common (e.g., \citealt{Arca-Sedda2019}).

A challenge in this census is that nuclear star clusters (and hence stripped nuclei) overlap in size and mass with globular clusters, which are far more numerous. A small number of the most massive globular clusters in the Milky Way and M31 have been identified as likely stripped nuclei \citep{Sarajedini1995,Hilker2000,Meylan2001,Pfeffer2021}, and in two cases have been shown to contain central black holes. The M31 globular cluster B023-G078, which has a stellar mass $\sim 6 \times 10^{6} M_{\odot}$, hosts a dynamically-detected $\sim 10^5 M_{\odot}$ supermassive black hole \citep{Pechetti2022}. $\omega$~Cen, the most massive Milky Way cluster ($4 \times 10^6 M_{\odot}$), has a recent {claimed} detection of an intermediate-mass black hole $> 8200 M_{\odot}$ via proper motions of individual fast stars at the cluster center \citep{Haberle2024}.

For galaxies beyond a few Mpc, due to the small angular sizes of the expected black hole spheres of influence, it is difficult or impossible to dynamically confirm black holes in the region of stellar mass overlap between UCDs and globular clusters. This has motivated efforts to search for alternative routes to uncover the presence of central black holes in stripped nuclei. 

Perhaps the most popular way to identify black holes in massive star clusters has been to search for multi-wavelength accretion signatures. The serendipitous X-ray detection of a tidal disruption event in a distant UCD provides compelling evidence for a central black hole \citep{Lin2018,Lin2020}, and the $L_X \gtrsim 10^{40}$ erg s$^{-1}$ source HLX-1 is plausibly a stripped nucleus with a central black hole as well \citep{Soria2017}, but such extreme events are rare. However, given the presence of ambient gas shed from evolving stars, central black holes in stripped nuclei should accrete some of this gas and glow in radio and X-ray as low-luminosity active galactic nuclei \citep{Maccarone2004}.

Radio continuum searches of massive star clusters in several nearby galaxies have not yet turned up any convincing candidates \citep{Wrobel2016,Wrobel2020}. While X-ray sources 
have been detected in a number of UCDs, in nearly all cases the X-ray emission can be plausibly explained by low-mass X-ray binaries. Indeed, the occurrence of X-ray sources in UCDs is lower than expected based on extrapolation of trends from lower-mass globular clusters \citep{Dabringhausen2012,Phillipps2013,Pandya2016,2016ApJ...819..164H}, suggesting little or no contribution from central black hole accretion.

Here we take a different tack. \citet{2022ApJ...929..147D} presented a large high-resolution spectroscopic survey of luminous globular clusters and candidate stripped nuclei in NGC 5128 (Cen A), with a goal of obtaining a complete sample of objects with $L_V \gtrsim 5 \times 10^{5} L_{\odot}$ (corresponding to $\sim 8 \times 10^5 M_{\odot}$) within a projected radius of 150 kpc of the galaxy center. This survey targeted candidates selected by \citet{2020ApJ...899..140V} and \citet{2021ApJ...914...16H} using photometry and structural information from both Gaia and ground-based imaging. When combined with reliable archival measurements, this effort resulted in a sample of 65 objects with measured dynamical mass-to-light ratios ($M/L_V$). The central result of \citet{2022ApJ...929..147D} was evidence for bimodality in $M/L_V$, with one subpopulation peaking around $\sim 1.3$ and the other at $\sim 2.7$. They argued that a natural explanation for this bimodality is if the high-$M/L_V$ group consists of stripped nuclei with embedded central black holes that make up typically $\sim 10\%$ of the remaining stellar mass of the former nucleus \citep{Mieske2013}. The low-$M/L_V$ group would then be either normal globular clusters or stripped nuclei that either lacked central black holes entirely, or had black holes of too low a mass ($\lesssim 10^5 M_{\odot}$) to have a detectable impact on the integrated velocity dispersion.

In this paper we analyze new and archival X-ray observations for this sample. This is the first direct comparison of X-ray properties for samples of dense star clusters where the identification of candidate stripped nuclei has been done using dynamical $M/L_V$ information rather than simply stellar mass. In principle, this should be more closely linked to the existence of a central black hole than stellar mass alone. In addition, the closer distance of Cen A (3.8 Mpc) compared to the Virgo or Fornax Clusters means that we are sensitive to lower X-ray luminosities than in most previous work systematically searching for X-ray emission from UCDs.

The paper is organized as follows.  In Section 2, we describe our  methods and data analysis for both {Chandra} and {XMM-Newton} X-ray data.  In Section 3, we present our X-ray luminosity measurements and assess them in the context of the existing $M/L_V$ information. Section 4 contains a discussion and summary of the results.

\section{X-ray Observations and Analysis}

Our paper is focused on the sample of 65 luminous Cen A globular clusters or stripped nuclei with dynamical $M/L_V$ \citep{2022ApJ...929..147D}, discussed in Section \ref{sec:intro}. 

\subsection{Chandra Data}

The majority of the sample had existing Chandra or XMM-Newton data.  For eight of the remaining objects, we obtained new Chandra observations (Proposal 23620148, P.I. Strader), using either ACIS-S (7\,ksec exposure time) or ACIS-I (9\,ksec exposure time), depending on the spatial distribution of the candidates.

The archival data is heterogeneous and in some cases involves multiple independent datasets covering the same source with a mixture of long and short exposure times.
For the sources with the largest number of observations and in turn the longest total exposure time, we stacked all ACIS observations above 30 ksec, as long as they were taken with the same chip (ACIS-S or ACIS-I). For sources with lower exposure time we stacked all observations regardless of exposure time or chip.

For the {Chandra} ACIS observations, we retrieved the data from the archive and re-processed it using the Chandra Interactive Analysis of Observations (\verb|CIAO|) version 4.16.0 \citep{2006SPIE.6270E..1VF}.  For each source, we defined circular extraction regions appropriately large to fully contain the source (dependent on the distance from the aim point).  We then extracted a local source-free background using circular regions that total three times the size of the target region.  

For the 16 sources with $\gtrsim50$ counts, we used the \verb|CIAO| task \verb|SPECEXTRACT| to extract a spectrum, using background and response files for each source and observation. We used XSPEC version 12.14.0h for spectral fitting \citep{1996ASPC..101...17A}, using a Tuebingen-Boulder ISM absorption model \citep{2000ApJ...542..914W}, \textit{tbabs}, and a power-law model. Spectra with counts $\gtrsim250$ were grouped to at least 15 counts per bin to allow $\chi^2$ statistics to be used. Spectra with fewer counts were left unbinned and fitted using Cash statistics \citep{1979ApJ...228..939C}. In both cases we performed fits with both $N_H$ free and fixed to the Galactic line of sight contribution ($2.4\times10^{20}\,$cm$^{-2}$; \citealt{2016A&A...594A.116H}). For sources with too few counts to perform robust spectral fitting, we measure the count rate, and then determine the unabsorbed flux assuming an absorbed power-law with $N_H$ fixed to the galactic line of sight contribution and fixed photon index $\Gamma=1.7$, a typical value expected for the sources in our sample. For all sources the flux reported is the unabsorbed 1--10\,keV flux.

In the case of non-detections, we used upper limits from the Chandra Source Catalog \citep{2010ApJS..189...37E}, using the flux\_sens\_b value and converting it to 1--10\,keV. For the handful of new observations that did not yet have upper limits listed in the Chandra Source Catalog, we calculated $2\sigma$ upper limits by {using the methodology of \citet{1991ApJ...374..344K} for a zero count background}, appropriate for short exposure times at these distant locations in the  Cen A halo.

Unabsorbed fluxes were converted to luminosities using a distance of 3.82\,Mpc to Cen A \citep{2010PASA...27..457H}. 
1--10 keV fluxes and corresponding luminosities for all sources are reported in Table \ref{table:main}. The spectral fit properties of sources with sufficient counts for spectral fitting are given in Table \ref{table:fits}. For sources with spectral fits, we use the fluxes from the free-$N_H$ fits in all of the other analysis and plots, but also note that the fluxes are typically consistent within 10\%, and none of our conclusions would be changed if instead we used the fixed-$N_H$ fits.

Some of the sources with fits in Table \ref{table:fits} show $N_H$ above that expected for Galactic foreground. In a few cases they sit spatially on the dust lane of Cen A which likely explains the inflated $N_H$: T17-1511 is a particularly clear example where the inferred $N_H$ of $5 \times 10^{21}$ cm$^{-2}$ is due to Cen A dust. In other cases, the excess $N_H$ could be due to small-scale less obvious dust, or could be internal to the X-ray source due e.g. to a more edge-on orientation.

\begin{figure*}[ht]
\centering
    \includegraphics[scale=0.24]{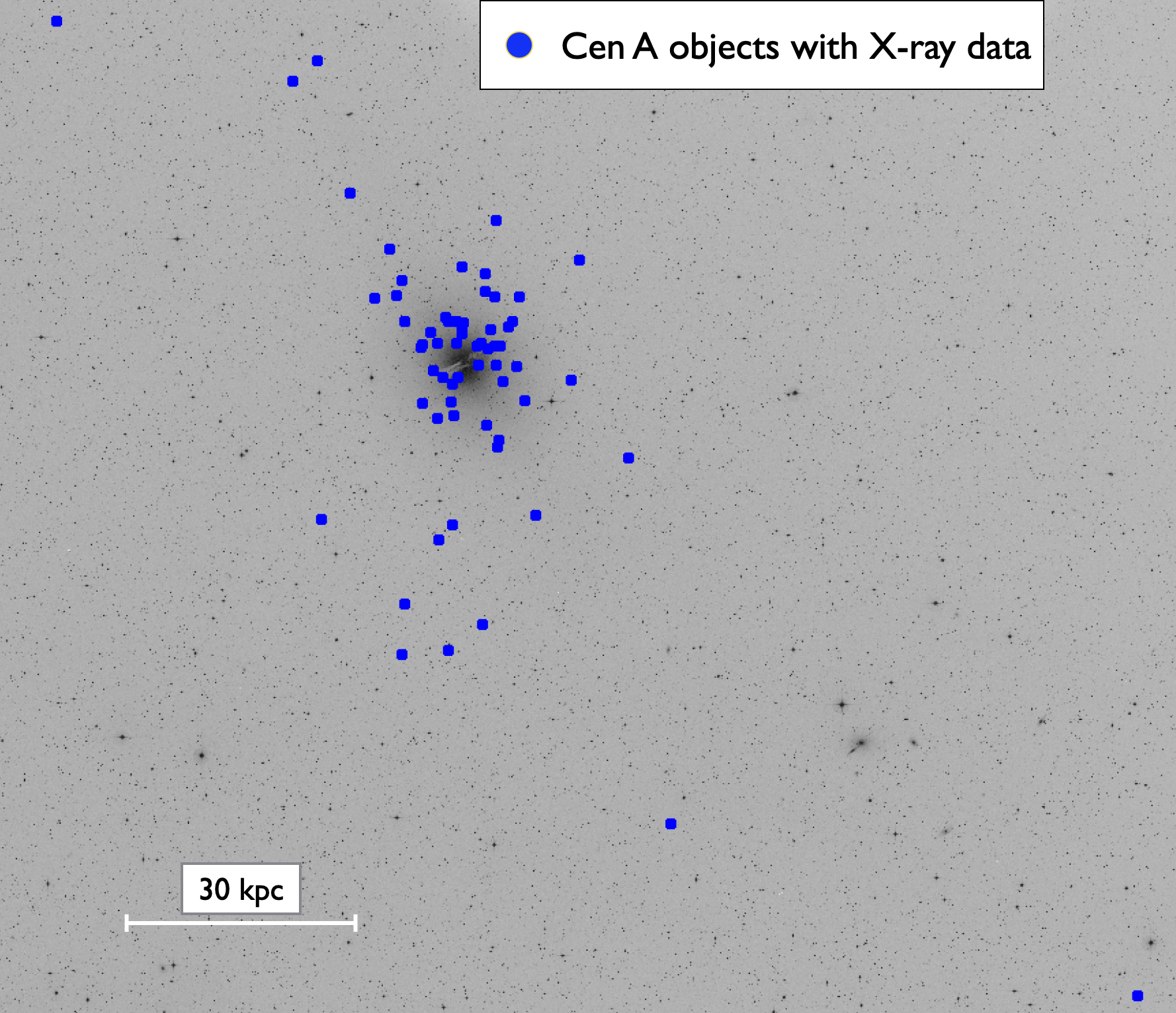}
    \caption{The 63 clusters in our sample plotted on a Digitized Sky Survey image of Cen A, with north up and east to the left. While many clusters are projected within the central 10s of kpc of the galaxy, some are in the distant outer halo, with the southernmost object at 116 kpc projected. The 30 kpc scale bar corresponds to $0.45^{\circ}$.}
    \label{fig:spat}
\end{figure*}

\subsection{XMM-Newton Data}

For 7 sources with no or limited ($< 10$ ksec) archival Chandra data, we supplement using publicly-available {XMM-Newton}/EPIC data. 
We processed the  data with the Science Analysis System (\verb|SAS|) version 1.3 (\verb|xmmsas_20230412_1735-21.0.0|). Similar to the {Chandra} observations, we defined circular regions around sources, and rectangular regions on occasions when sources were near chip gaps.  We extracted the background using nearby regions at least three times larger than the source region. We selected single and double events (pattern 0-4 for pn and 0-12 for MOS) with standard flagging criteria \verb|#XMMEA_EP| for pn or \verb|#XMMEA_EM| for MOS, in addition to \verb|FLAG=0|.  We extracted individual MOS and pn spectra using standard \textit{xmmselect} tasks, before combing them with \textit{epicspeccombine}. The spectral fitting followed the same procedures as for the Chandra data.

For undetected sources, {XMM} flux upper limits were obtained from the RapidXMM upper limit server \citep{2022MNRAS.511.4265R}, using the 4XMM catalog \citep{2020A&A...641A.136W}. Wherever data from multiple cameras (MOS1, MOS2, and pn) were available for the same location, we stacked the limiting sensitivities to produce a deeper merged EPIC upper limit.

The results of the XMM analysis are also listed in Tables \ref{table:main} and  \ref{table:fits}.

\subsection{Final Sample Size}

Of the initial sample of 65 clusters, 63 had an X-ray detection or constraining upper limit from either Chandra or XMM. For the remaining two objects, HGHH-41 fell on a Chandra chip gap, and H12\_141 had an XMM slew observation that was too shallow to be useful. These objects are listed at the end of Table \ref{table:main} for completeness but not included in any subsequent analysis.

We emphasize that most of the 21 X-ray detections have been previously published \citep{Kraft2001,Minniti2004,Woodley2008,Zhang2011,Pandya2016}, with the exceptions being HH-22, HHH86-34, and HGHH-G251. The central novelty here is the systemic analysis of a dynamically sorted sample of objects.

Figure \ref{fig:spat} plots the 63 targets with X-ray data on an image of Cen A, showing they extend to the far outer halo.

\section{Results}

\subsection{$M/L_V$ Subsamples}

The main goal of the study was to compare X-ray properties for the high and low-$M/L_V$ subsamples of the 63 sources with X-ray data. We adopt the \citet{2022ApJ...929..147D} division of the sample at $M/L_V = 2.3$, which puts 41 sources in the low-$M/L_V$ group and 22 sources in the high-$M/L_V$ group. The number of X-ray detections in the groups are 16/41 ($39^{+8}_{-7}$\%) and 5/22 ($23^{+11}_{-6}$\%), respectively, where the uncertainties are $1\sigma$ equivalent Bayesian confidence intervals for the binomial distribution following the methodology of \citet{Cameron2011}. 
The median X-ray upper limit for the non-detected sources is nearly identical for the two groups ($\sim 5 \times 10^{36}$ erg s$^{-1}$), implying that the limits are similarly sensitive despite the heterogeneous nature of the sample. The rates of X-ray detection between the subsamples differ only at the $1.5\sigma$ level, implying that there is no evidence that the high-$M/L_V$ candidate stripped nuclei host X-ray sources at a rate higher than normal globular clusters.
The result is the same if restricted to the more luminous clusters: using only those with $M_V < -10$, 2/13 ($15^{+15}_{-5}$\%) of the high-$M/L_V$ group and 8/21 ($38^{+11}_{-9}$\%) of the low-$M/L_V$ group are detected in X-rays.

\begin{figure}
    \centering
    \includegraphics[scale=0.58]{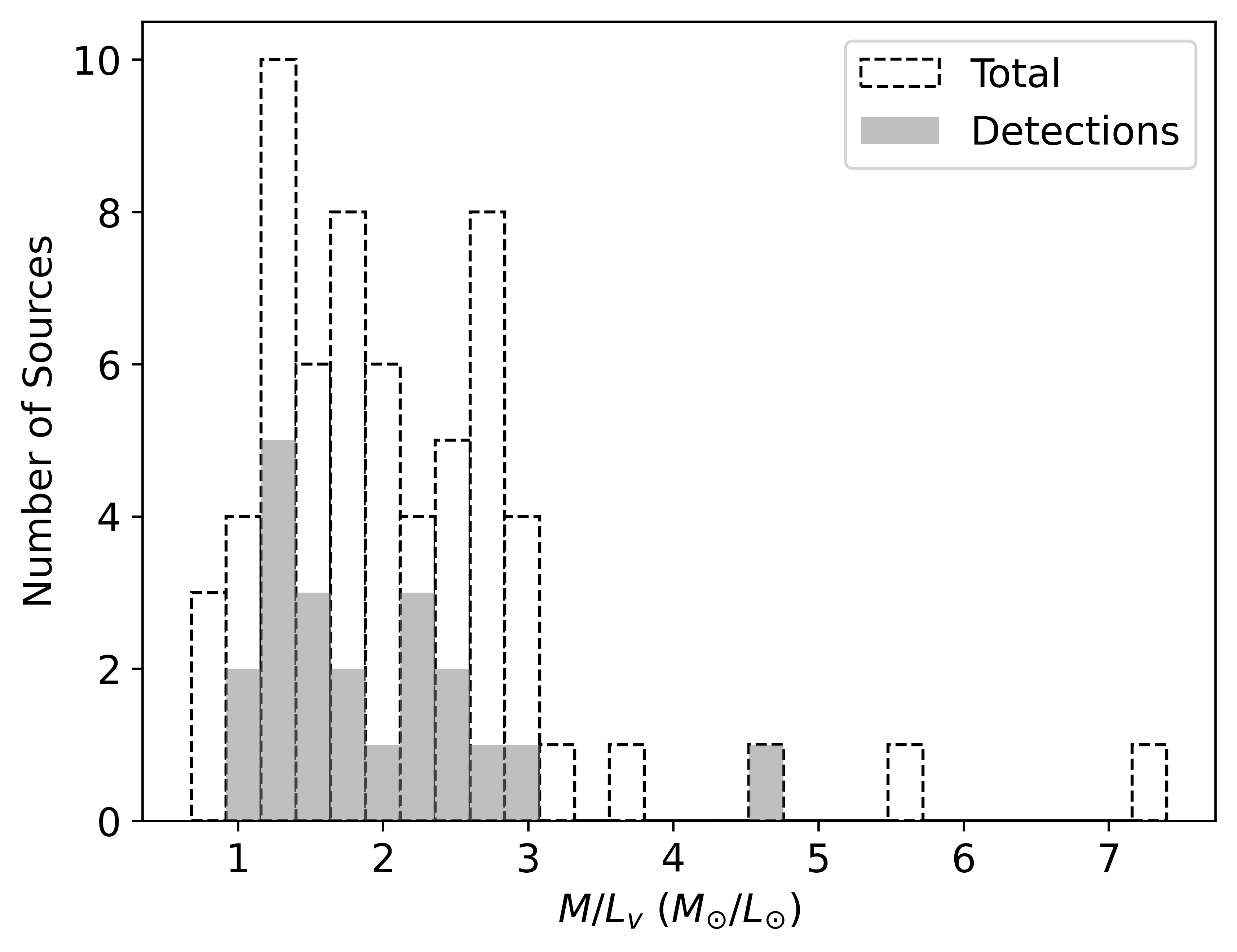}
    \caption{Histogram of $M/L_V$ for all 63 objects (dashed line) and X-ray detections (shaded). The bimodal distribution of $M/L_V$ is visible in both samples.}
    \label{fig:ml_hist}
\end{figure}

\begin{figure}
    \centering
    \includegraphics[scale=0.58]{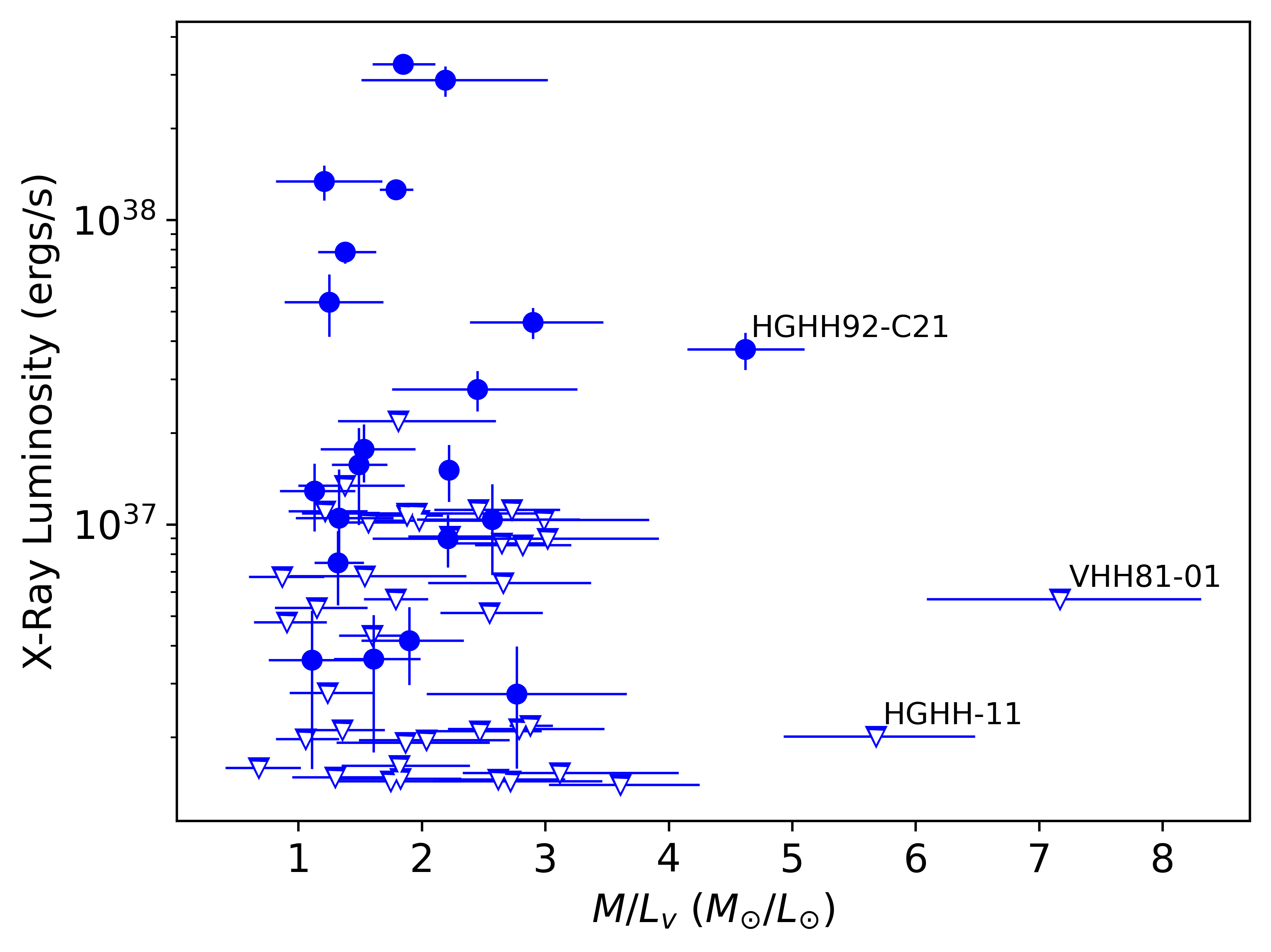}
    \caption{1--10 keV X-ray luminosity vs.~$M/L_V$ for all sources, with detections shown as filled circled and non-detections (upper limits) shown as unfilled downward triangles. The similarity of the X-ray properties of the low and high-$M/L_V$ objects is evident. There are three sources with $M/L_V$ much larger than the other clusters, which have especially strong dynamical cases for being stripped nuclei.} 
        \label{fig:ml_plot}
\end{figure}

The central result is shown in Figures \ref{fig:ml_hist} and \ref{fig:ml_plot}. Figure \ref{fig:ml_hist} shows a histogram of the 
$M/L_V$ measurements with the X-ray detections shaded, and Figure \ref{fig:ml_plot} plots the X-ray luminosity or upper limit against  $M/L_V$. In both figures the similarity of the X-ray properties of the high ($M/L_V > 2.3$) and low-$M/L_V$ ($< 2.3$) groups is evident.
The three objects with very high 
$M/L_V$ ($> 4$) are discussed individually in Section 3.4.

\subsection{Eddington Ratio Limits for Candidate Stripped Nuclei}

Of the 22 clusters with high $M/L_V$, 18 have mass estimates for a central black hole from \citet{2022ApJ...929..147D}, and of these 18, 4 have X-ray detections. These were estimated under the assumption that the underlying stellar population has a $M/L_V$ equal to the mean of the low-$M/L_V$ sample ($M/L_V = 1.51$) and then a dynamical model was created for each that added a black hole of the appropriate mass to reproduce the higher observed $M/L_V$. While these are necessarily uncertain on an object-by-object basis, and require future high spatial resolution integral field spectroscopy to confirm the dynamical presence of a central black hole, these masses should be on average reasonable if central black holes are the explanation for the elevated $M/L_V$ for these objects. The median inferred central black hole mass is $2 \times 10^{5} M_{\odot}$.

Under these assumptions, we can calculate upper limits to the Eddington fraction---the ratio of the active galactic nucleus bolometric luminosity ($L_{bol}$) to the Eddington luminosity ($L_{edd}$)---for the putative central black holes in the 14 sources that are \emph{undetected} in the X-rays. To do this, we convert the 1--10 keV X-ray luminosity upper limit to a 2--10 keV limit assuming a power-law with $\Gamma = 1.7$, and then take the bolometric correction measured for low-luminosity active galactic nuclei, $L_{bol} = 15.8 \, L_X$ (2--10 keV), from \citet{Ho2009}. The resulting Eddington ratio upper limits are listed in Table \ref{table:main}. 

The median upper limit is $L_{bol}/L_{edd} < 2 \times 10^{-6}$, with the strongest upper limits about a factor of 3 lower than this. The median $L_{bol}/L_{edd}$ of \emph{detected} low-luminosity active galactic nuclei in the \citet{Ho2009} sample is somewhat higher, at $L_{bol}/L_{edd} \sim 5 \times 10^{-6}$. Nearly all these galaxies are much more massive than the predicted former host galaxies of the stripped nuclei under consideration here, let alone the stripped nuclei themselves. Nonetheless, it does suggest preliminary evidence that if high-$M/L_V$ stripped nuclei mostly contain central black holes, the Eddington ratio distribution for these sources is different than for low-luminosity active galactic nuclei in more massive galaxies.
This is given further weight by the strict upper limits on the Eddington ratios of the dynamically detected black holes in the M31 stripped nucleus B023-G078 ($L_{bol}/L_{edd} \lesssim 2 \times 10^{-7}$; \citealt{Pechetti2022}) and in $\omega$ Cen ($L_{bol}/L_{edd} \lesssim 10^{-12}$; \citealt{Haberle2024}).

{This Eddington ratio analysis is agnostic as to the physical cause of the faintness of the black holes. By making additional assumptions, it is possible to tie the results more closely to potential physical parameters of interest in the context of a specific model. The Bondi accretion rate $\dot{M}_{bondi}$ (in units of g s$^{-1}$) for a black hole accreting isothermal ionized gas with $T=10^4$ K and a mean molecular mass $\mu = 0.59$ (reasonable for a typical star cluster) is
$\dot{M}_{bondi} = 1.48 \times 10^{11}\,M^2\,n$, where $M$ is the mass in solar masses and $n$ is the gas density (see \citealt{Tremou2018}). The observed X-ray luminosity is given by $L_X = \epsilon \dot{M}_{bondi} c^2$ where $\epsilon$ is the radiative efficiency. We further assume radiatively inefficient accretion (e.g., \citealt{Narayan1995}), such that $\epsilon$ scales with the accretion rate (e.g., \citealt{Merloni2003}) of a form used by previous works that gives continuity at the hard/soft transition for black hole accretion:  $\epsilon = 0.1(\dot{M}_{bondi}/\dot{M}_{edd})/0.02$ \citep{Maccarone2005,Maccarone2008,Strader2012}. Combining these expressions gives $L_X = 7.0 \times 10^{25} \, M^3 \, n^2 $. We can then use the X-ray upper limits or detections for the 18 Cen A objects with central black hole mass estimates to determine upper limits or estimates of the central gas density, assuming radiatively inefficient Bondi accretion as outlined here. These values are listed in Table \ref{table:main}. The median upper limit is $n < 0.002$ cm$^{-3}$, which is about a factor of 100 lower than inferred in the core of the globular cluster 47 Tuc, where timing of a population of millisecond pulsars has allowed a measurement of the gas density $n = 0.23\pm0.05$ cm$^{-3}$ \citep{Abbate2018}. If we interpret the four X-ray detections with the same model, the inferred gas densities are $n = 0.005-0.01$ cm$^{-3}$, still extremely low. 

Alternatively, if we assume a gas density as observed in 47 Tuc ($n \sim 0.2$ cm$^{-3}$), then the X-ray measurements would imply accretion at a lower rate than Bondi and/or at lower radiative efficiency than assumed. If the accretion rate was a fraction $f$ of the Bondi rate, then
in the above framework $L_X \propto f^2$, and the X-ray upper limits would correspond to $f \sim 0.002$--0.1, depending on the cluster. If instead the accretion was at the Bondi rate, the X-ray upper limits would imply radiative efficiencies in the range $\epsilon \lesssim 3\times 10^{-7}$--$10^{-5}$, orders of magnitude lower than expected for Bondi accretion in the radiatively inefficient formalism assumed above. A combination of sub-Bondi accretion and lower radiative efficiency is also possible. Similar results have be found for central black holes in some quiescent early-type galaxies, which also show accretion below the expected Bondi rate and/or very low radiative efficiency \citep{Pellegrini2005,Soria2006,Balmaverde2008,Ho2009}.

\subsection{Interpreting the X-ray Detections}

Because most of the candidate stripped nuclei were not detected in X-rays, the above discussion is largely independent of the interpretation of the 4 detected X-ray sources in the high-$M/L_V$ clusters. Given the lower fraction of X-ray detections among high-$M/L_V$ clusters compared to low-$M/L_V$ clusters, it is a priori plausible that all of these detections are associated with low-mass X-ray binaries rather than with active galactic nuclei. The one exception is HGHH92-C21, which we discuss in more detail in the next subsection.

\subsection{The Highest $M/L_V$ Objects}

\subsubsection{HGHH92-C21}

One of the four high-$M/L_V$ sources that is detected in X-rays (mean $L_X \sim 4 \times 10^{37}$ erg s$^{-1}$) is HGHH92-C21. This cluster has the third-highest $M/L_V$ in our sample ($4.6\pm0.5$), far higher than observed for typical globular clusters and consistent with an elevated velocity dispersion due to a central massive black hole. The cluster has a large half-mass radius ($\sim 9.2$ pc) and is very flattened ($\epsilon = 0.33$; \citealt{Harris2002}), a characteristic shared by nuclear star clusters (e.g., \citealt{Seth2006}). \citet{Voggel2018} obtained ground-based adaptive-optics assisted integral field spectroscopy to search for a central black hole in HGHH92-C21, but owing to poor data quality were only able to obtain an upper limit of $< 10^6 M_{\odot}$ on the mass of a black hole.

The X-ray detection of this cluster was previously noted in the Chandra imaging of \citet{Kraft2001}. Subsequent work showed that the X-ray source shows frequent short timescale (hundreds of seconds) flares up to $\sim 10^{40}$ erg s$^{-1}$, a factor of $\sim 200$--300 higher than its mean luminosity \citep{2016Natur.538..356I}. While a stellar-mass compact object cannot be definitively excluded as the source of these luminous flares, the X-ray luminosities and flare timescales would be easily accommodated by a $\lesssim 10^6 M_{\odot}$ central black hole. HGHH92-C21 is therefore the object in our sample in which the X-ray emission is most likely to have an origin in an active galactic nucleus. Radio continuum imaging to further test this interpretation would be valuable.

\subsubsection{VHH81-01 and HGHH-11}

VHH81-01 and HGHH-11 stand out as having the highest $M/L_V$ in our sample: $7.2\pm1.1$ and $5.7\pm0.8$, respectively, making them strong candidate stripped nuclei with central black holes. This is especially true for VHH81-01 which has a remarkably large half-mass radius of 31.5 pc \citep{2022ApJ...929..147D}.

Neither is X-ray detected. Despite the relatively shallow XMM data available for VHH81-01, because of its high inferred black hole mass ($8.6^{+10.3}_{-0.1} \times 10^{5} M_{\odot}$; \citealt{{2022ApJ...929..147D}}), the Eddington limit is the lowest of any cluster in our sample at $< 6 \times 10^{-7}$. Hence new, deeper X-ray observations could allow even tighter constraints on the Eddington fraction of the central black hole likely present.

\section{Discussion and Conclusions}

We have presented X-ray constraints or detections for a sample of 63 luminous globular clusters or stripped nuclei in Cen A, all of which have dynamical mass-to-light measurements.

Our main finding is that there is no enhancement of X-ray sources in luminous star clusters in Cen A that show dynamical evidence for hosting a massive central black hole---likely because most are stripped nuclei---compared to a control sample of normal luminous globular clusters. This result is the straightforward one expected from an extrapolation of previous work, which found that candidate stripped nuclei host X-ray sources at a rate at most comparable to, and usually lower than, other massive globular clusters \citep{Dabringhausen2012,Phillipps2013,Pandya2016}. 

There are two ways in which our new result is consistent with but has broader implications than previous work. The first is the sample selection: rather than identifying candidate stripped nuclei solely by mass or size, it is done here by dynamical $M/L_V$, and the two subsamples have similar stellar masses and sizes, excepting a few outliers. The dynamical selection of the candidate stripped nuclei should enhance the purity of the sample (i.e., minimize contamination from massive globular clusters), and the rate of dynamically formed X-ray binaries ought to be similar between the subsamples, allowing a direct comparison between them. 

The other difference compared to past papers is the use of deep X-ray data for a relatively nearby galaxy. The median X-ray upper limit of $5 \times 10^{36}$ erg s$^{-1}$ is more than an order of magnitude deeper than in the study of \citet{Pandya2016}, which used a heterogeneous sample of galaxies with a range of environments and distances, mostly at or beyond the Virgo Cluster. This is why we see a much higher X-ray detection rate (23\% for the high-$M/L_V$ subset and 33\% for the full sample) than the $3\%$ detection rate reported by that previous paper. It also means that we can reach   Eddington ratios expected for low-luminosity active galactic nuclei, such that the X-ray limits are constraining at the expected black hole masses.

Despite these much deeper X-ray limits, we still find no evidence for an excess of X-ray sources in candidate stripped nuclei. If the elevated $M/L_V$ observed in these objects is indeed due to massive black holes in many or most cases, then they must be typically accreting at $L_{bol}/L_{edd} < 2 \times 10^{-6}$. The recently discovered intermediate-mass black hole in the stripped nucleus $\omega$ Cen is remarkably faint ($L_{bol}/L_{edd} \lesssim 10^{-12}$; \citealt{Haberle2024}), and there is some evidence that central black holes in some nearby low-mass early-type galaxies may also have relatively low Eddington fractions \citep{Urquhart2022}. Together these observations represent emerging evidence that massive black holes in stripped nuclei---and perhaps also at the centers of low-mass galaxies that are not actively forming stars---accrete at low rates that will make them difficult to discover and characterize using X-ray or radio observations.

{An alternative interpretation of these results is that there is no enhancement in X-ray sources in the high-$M/L_V$ objects because the latter do not contain central black holes. This would, in turn, imply a low black hole occupation fraction for low-mass galaxies. In this case an alternative explanation for the objects with elevated $M/L_V$ is needed. It is unlikely to be either metallicity or age: \citet{2022ApJ...929..147D} show there is no observed correlation between $M/L_V$ and metallicity for the Cen A sample. While precise ages are not available for these objects, the low-$M/L_V$ clusters have $M/L_V$ values similar to old ($\gtrsim 10$ Gyr) Galactic globular clusters, consistent with similar ages. At old ages, even large age differences do not produce large differences in $M/L$: for example, using a Kroupa initial mass function and a metallicity of [$Z$/H] = --1.0, the simple stellar population models of \citet{Conroy2010} imply only a 20\% increase in $M/L_V$ from 10 to 13 Gyr, far less than the factor of $\sim 2$ difference observed between the low and high-$M/L_V$ populations. Alternatively, one could argue that the low-$M/L_V$ population is instead associated with an intermediate-age stellar population known to be present in Cen A, perhaps due to a recent merger (e.g., \citealt{Rejkuba2001}). But this explanation is inconsistent both with the large metallicity range for the low-$M/L_V$ population and for the similar spatial distributions of the low and high-$M/L_V$ populations, which have comparable  median projected galactocentric radii \citep{2022ApJ...929..147D}. This latter point also disfavors tidal disequilibrium (e.g., \citealt{Forbes2014}) as an explanation, since tidal forces are strongly dependent on galactocentric radius.


A variation in the initial mass function, with either an enhancement of M dwarfs or of high-mass stars that leave compact remnants \citep{Dabringhausen2012}, could also lead to elevated $M/L$. As pointed out by \citet{2022ApJ...929..147D}, in the UCDs with radial $M/L$ profiles, the $M/L$ is centrally peaked in a manner inconsistent with an enhancement of low-mass stars, which would have too long a mass segregation timescale in these massive clusters. Stellar-mass black holes can mass segregate on a much shorter timescale, so in principle, an unexpectedly large population of stellar-mass black holes could explain the elevated $M/L$ in some clusters. This does not seem to naturally explain the bimodal distribution of $M/L$ nor the lack of dependence on metallicity, but cannot be ruled out with existing data for most of the Cen A objects.}


To take the next step will require more precise dynamical measurements of the suspected central black holes in many of these objects, which will allow their confirmation and improved constraints on their accretion properties. Unfortunately, such measurements are challenging with currently available instrumentation \citep{Voggel2018}, and accumulating a sufficient sample of precise measurements may require 30-m telescopes \citep{Do2014,Nguyen2024}.

\begin{acknowledgments}

We acknowledge the helpful comments of an anonymous referee.

Support for this work was provided by the National Aeronautics and Space Administration through Chandra Award Number GO2-23061X issued by the Chandra X-ray Observatory Center, which is operated by the Smithsonian Astrophysical Observatory for and on behalf of the National Aeronautics Space Administration under contract NAS8-03060. We acknowledge support from NASA grant 80NSSC21K0628. This work was supported by the NSF REU program, NSF Division of Physics, Award No. 2050733. We acknowledge support from the Packard Foundation.

This research has made use of data obtained from the Chandra Data Archive and the Chandra Source Catalog, both provided by the Chandra X-ray Center (CXC).  This paper also employs a list of Chandra datasets, obtained by the Chandra X-ray Observatory, contained in the Chandra Data Collection (CDC) ~\dataset[doi:10.25574/cdc.348]{https://doi.org/10.25574/cdc.348}

Based on observations obtained with XMM-Newton, an ESA science mission with instruments and contributions directly funded by ESA Member States and NASA.  

\end{acknowledgments}

\bibliography{references}{}
\bibliographystyle{aasjournal}

\clearpage
\startlongtable
\tablewidth{0pt}
\rotate
\begin{deluxetable*}{lllccllllll}
\label{table:main}
\tablecaption{X-Ray Properties of Target Clusters}
\centering
\tabletypesize{\scriptsize}
\tablehead{
\colhead{ID\tablenotemark{a}} & \colhead{R.A. (J2000)} & \colhead{Decl. (J2000)} & \colhead{$M_V$\tablenotemark{b}} & \colhead{M/$L_V$\tablenotemark{c}} & \colhead{X-ray flux\tablenotemark{d}} & \colhead{X-Ray lum.\tablenotemark{e}} & \colhead{BH Mass\tablenotemark{f}} & \colhead{Edd. Fract.\tablenotemark{g}} & \colhead{Gas $n$ \tablenotemark{h}} & \colhead{Instr.\tablenotemark{i}}\\
\colhead{ } & \colhead{Degrees} & \colhead{Degrees} & \colhead{(mag)} & \colhead{$M_{\odot}$/$L_{\odot}$} & \colhead{$10^{-15}$ ergs s$^{-1}$} & \colhead{$10^{36}$ ergs s$^{-1}$} & \colhead{$10^5$ $M_{\odot}$} & \colhead{$10^{-6}$}  & \colhead{$10^{-3}$ cm$^{-3}$}  & \colhead{}   }
\startdata
HGHH92-C23 & $201.477417$ & $-42.990389$ & $-11.66$ & $1.79^{+0.14}_{-0.13}$ & $72.1^{+4.0}_{-4.1}$ & $126.0^{+7.0}_{-7.2}$ & \nodata & \nodata & \nodata & ACIS-I \\ 
HGHH-07 & $201.522474$ & $-42.942327$ & $-11.09$ & $2.22^{+0.07}_{-0.07}$ & $8.5^{+1.6}_{-1.6}$ & $14.8^{+2.7}_{-2.7}$ & \nodata & \nodata & \nodata & ACIS-I \\ 
HHH86-30 & $201.226440$ & $-42.890201$ & $-11.02$ & $1.54^{+0.82}_{-0.61}$ & $<3.9^{ }_{ }$ & $<6.8^{ }_{ }$ & \nodata & \nodata & \nodata & ACIS \\ 
HH-10 & $201.379309$ & $-42.837526$ & $-10.98$ & $2.66^{+0.71}_{-0.61}$ & $<3.7^{ }_{ }$ & $<6.4^{ }_{ }$ & $3.2^{+6.6}_{-1.7}$ & $<1.9^{ }_{ }$ & $<1.7$ & ACIS \\ 
K-029 & $201.288256$ & $-42.983105$ & $-10.89$ & $1.38^{+0.25}_{-0.22}$ & $44.9^{+3.6}_{-3.7}$ & $78.5^{+6.2}_{-6.5}$ & \nodata & \nodata & \nodata & ACIS-I \\ 
VHH81-01 & $200.934030$ & $-43.186620$ & $-10.80$ & $7.17^{+1.14}_{-1.08}$ & $<3.3^{ }_{ }$ & $<5.7^{ }_{ }$ & $8.6^{+10.0}_{-0.1}$ & $<0.6^{ }_{ }$ & $<0.4$ & EPIC \\ 
VHH81-03 & $201.242499$ & $-42.936124$ & $-10.65$ & $1.32^{+0.21}_{-0.19}$ & $4.3^{+1.2}_{-1.2}$ & $7.5^{+2.0}_{-2.0}$ & \nodata & \nodata & \nodata & ACIS-I \\ 
VHH81-5 & $201.317123$ & $-42.882801$ & $-10.63$ & $2.55^{+0.43}_{-0.40}$ & $<2.9^{ }_{ }$ & $<5.1^{ }_{ }$ & $2.0^{+4.1}_{-0.5}$ & $<2.5^{ }_{ }$ & $<3.0$ & ACIS \\ 
HGHH92-C17 & $201.415542$ & $-42.933111$ & $-10.63$ & $3.61^{+0.64}_{-0.58}$ & $<0.8^{ }_{ }$ & $<1.4^{ }_{ }$ & \nodata & \nodata & \nodata & ACIS \\ 
KV19-442 & $202.432394$ & $-42.391404$ & $-10.55$ & $1.57^{+0.31}_{-0.28}$ & $<5.8^{ }_{ }$ & $<10^{ }_{ }$ & \nodata & \nodata & \nodata & ACIS-S \\ 
K-034 & $201.292745$ & $-42.892504$ & $-10.53$ & $1.21^{+0.47}_{-0.39}$ & $77^{+10}_{-10}$ & $134^{+17}_{-18}$ & \nodata & \nodata & \nodata & ACIS-I \\ 
KV19-271 & $201.296298$ & $-43.509212$ & $-10.41$ & $2.65^{+0.39}_{-0.38}$ & $<5.0^{ }_{ }$ & $<8.7^{ }_{ }$ & $4.5^{+2.8}_{-1.4}$ & $<1.9^{ }_{ }$ & $<1.2$ & ACIS-I \\ 
HGHH92-C21 & $201.469750$ & $-43.096222$ & $-10.39$ & $4.62^{+0.48}_{-0.47}$ & $21.5^{+2.9}_{-3.0}$ & $37.5^{+5.1}_{-5.4}$ & \nodata & \nodata & \nodata & ACIS-I \\ 
HHH86-14 & $201.293668$ & $-42.747977$ & $-10.39$ & $1.81^{+0.79}_{-0.49}$ & $<13^{ }_{ }$ & $<22^{ }_{ }$ & \nodata & \nodata & \nodata & ACIS \\ 
KV19-289 & $201.380317$ & $-43.046136$ & $-10.37$ & $1.87^{+0.68}_{-0.56}$ & $<1.1^{ }_{ }$ & $<1.9^{ }_{ }$ & \nodata & \nodata & \nodata & ACIS \\ 
HGHH-11 & $201.227938$ & $-43.022712$ & $-10.35$ & $5.68^{+0.80}_{-0.75}$ & $<1.2^{ }_{ }$ & $<2.0^{ }_{ }$ & \nodata & \nodata & \nodata & ACIS \\ 
HGHH-12 & $201.273697$ & $-43.175240$ & $-10.35$ & $1.49^{+0.23}_{-0.22}$ & $8.0^{+2.3}_{-2.2}$ & $14.0^{+4.0}_{-3.8}$ & \nodata & \nodata & \nodata & ACIS-I \\ 
KV19-212 & $200.790847$ & $-43.874458$ & $-10.33$ & $2.73^{+0.39}_{-0.39}$ & $<6.4^{ }_{ }$ & $<11^{ }_{ }$ & $3.5^{+2.8}_{-0.9}$ & $<3.1^{ }_{ }$ & $<1.9$ & ACIS-S \\ 
T17-1412 & $201.281782$ & $-43.020912$ & $-10.30$ & $1.85^{+0.26}_{-0.25}$ & $186^{+9}_{-9}$ & $325^{+16}_{-16}$ & \nodata & \nodata & \nodata & ACIS-I \\ 
HGHH-35 & $201.434161$ & $-42.983166$ & $-10.30$ & $1.36^{+0.34}_{-0.27}$ & $<1.2^{ }_{ }$ & $<2.1^{ }_{ }$ & \nodata & \nodata & \nodata & ACIS \\ 
HCH99 21 & $201.394375$ & $-43.057694$ & $-10.28$ & $0.68^{+0.34}_{-0.27}$ & $<0.9^{ }_{ }$ & $<1.6^{ }_{ }$ & \nodata & \nodata & \nodata & ACIS \\ 
H12$\_78$ & $201.672201$ & $-42.703936$ & $-10.25$ & $2.82^{+0.39}_{-0.39}$ & $<4.9^{ }_{ }$ & $<8.6^{ }_{ }$ & $4.6^{+2.4}_{-1.2}$ & $<1.8^{ }_{ }$ & $<1.1$ & ACIS \\ 
HHH86-29 & $201.168230$ & $-43.301482$ & $-10.25$ & $2.46^{+0.39}_{-0.36}$ & $<6.4^{ }_{ }$ & $<11^{ }_{ }$ & $1.6^{+3.7}_{-0.4}$ & $<6.7^{ }_{ }$ & $<6.2$ & ACIS-S \\ 
HH-22 & $201.089178$ & $-43.043633$ & $-10.18$ & $1.25^{+0.44}_{-0.36}$ & $85^{+13}_{-13}$ & $149^{+23}_{-23}$ & \nodata & \nodata & \nodata & ACIS-I \\ 
HHH86-26 & $201.563543$ & $-42.808168$ & $-10.12$ & $1.22^{+0.34}_{-0.30}$ & $<6.3^{ }_{ }$ & $<11^{ }_{ }$ & \nodata & \nodata & \nodata & ACIS \\ 
HGHH92-C22 & $201.473208$ & $-42.985444$ & $-10.11$ & $2.88^{+0.18}_{-0.17}$ & $<1.3^{ }_{ }$ & $<2.2^{ }_{ }$ & \nodata & \nodata & \nodata & ACIS \\ 
HGHH-G342 & $201.274196$ & $-42.983493$ & $-10.07$ & $1.30^{+0.42}_{-0.35}$ & $<0.9^{ }_{ }$ & $<1.5^{ }_{ }$ & \nodata & \nodata & \nodata & ACIS \\ 
T17-1511 & $201.327091$ & $-43.021130$ & $-10.06$ & $2.45^{+0.81}_{-0.69}$ & $15.9^{+2.4}_{-2.4}$ & $27.8^{+4.2}_{-4.3}$ & $2.4^{+3.4}_{-1.7}$ & $10^{+10}_{-7}$ & 5.4 & ACIS-I \\ 
HGHH-19 & $201.430768$ & $-43.123036$ & $-10.06$ & $1.24^{+0.37}_{-0.31}$ & $<1.6^{ }_{ }$ & $<2.8^{ }_{ }$ & \nodata & \nodata & \nodata & ACIS \\ 
R261 & $201.303750$ & $-43.133083$ & $-10.06$ & $1.06^{+0.27}_{-0.24}$ & $<1.1^{ }_{ }$ & $<2.0^{ }_{ }$ & \nodata & \nodata & \nodata & ACIS \\ 
H12$\_95$ & $201.500890$ & $-43.475899$ & $-10.05$ & $1.88^{+0.29}_{-0.27}$ & $<6.1^{ }_{ }$ & $<11^{ }_{ }$ & \nodata & \nodata & \nodata & EPIC \\ 
KV19-288 & $201.369036$ & $-42.941958$ & $-10.02$ & $2.72^{+0.74}_{-0.63}$ & $<0.8^{ }_{ }$ & $<1.4^{ }_{ }$ & $1.9^{+2.8}_{-0.7}$ & $<0.7^{ }_{ }$ & $<1.7$ & ACIS \\ 
AAT329848 & $201.505241$ & $-43.570995$ & $-9.95$ & $1.96^{+0.29}_{-0.29}$ & $<6.2^{ }_{ }$ & $<11^{ }_{ }$ & \nodata & \nodata & \nodata & ACIS-I \\ 
WHH-17 & $201.371848$ & $-42.963098$ & $-9.95$ & $1.90^{+0.44}_{-0.39}$ & $2.4^{+0.7}_{-0.7}$ & $4.2^{+1.2}_{-1.2}$ & \nodata & \nodata & \nodata & ACIS-I \\ 
H12$\_106$ & $201.386037$ & $-43.560646$ & $-9.93$ & $1.88^{+1.13}_{-0.85}$ & $<6.2^{ }_{ }$ & $<11^{ }_{ }$ & \nodata & \nodata & \nodata & ACIS-I \\ 
PFF-GC098 & $201.724630$ & $-43.321592$ & $-9.92$ & $1.79^{+0.26}_{-0.26}$ & $<3.3^{ }_{ }$ & $<5.7^{ }_{ }$ & \nodata & \nodata & \nodata & EPIC \\ 
HHH86-36 & $201.532145$ & $-42.866721$ & $-9.91$ & $2.21^{+0.74}_{-0.61}$ & $5.2^{+1.0}_{-1.0}$ & $9.0^{+1.8}_{-1.8}$ & \nodata & \nodata & \nodata & ACIS-I \\ 
PFF-GC100 & $201.764180$ & $-42.454757$ & $-9.86$ & $1.38^{+0.48}_{-0.38}$ & $<7.7^{ }_{ }$ & $<13^{ }_{ }$ & \nodata & \nodata & \nodata & EPIC \\ 
HGHH-G204 & $201.445728$ & $-43.034851$ & $-9.85$ & $2.62^{+0.54}_{-0.48}$ & $<0.8^{ }_{ }$ & $<1.5^{ }_{ }$ & $1.9^{+1.9}_{-0.8}$ & $<0.7^{ }_{ }$ & $<1.8$ & ACIS \\ 
HHH86-34 & $201.419139$ & $-43.353852$ & $-9.85$ & $2.19^{+0.83}_{-0.68}$ & $225^{+16}_{-17}$ & $393^{+28}_{-30}$ & \nodata & \nodata & \nodata & EPIC \\ 
WHH-22 & $201.397118$ & $-43.091414$ & $-9.83$ & $1.11^{+0.44}_{-0.35}$ & $1.9^{+0.3}_{-0.4}$ & $3.4^{+0.6}_{-0.7}$ & \nodata & \nodata & \nodata & ACIS-I \\ 
HHH86-37 & $201.544021$ & $-42.895178$ & $-9.83$ & $1.13^{+0.33}_{-0.28}$ & $7.7^{+1.9}_{-2.0}$ & $13.5^{+3.3}_{-3.5}$ & \nodata & \nodata & \nodata & ACIS-I \\ 
WHH-18 & $201.375285$ & $-42.946367$ & $-9.78$ & $1.83^{+0.49}_{-0.51}$ & $<0.8^{ }_{ }$ & $<1.5^{ }_{ }$ & \nodata & \nodata & \nodata & ACIS \\ 
HHH86-28 & $201.075206$ & $-42.816956$ & $-9.75$ & $1.98^{+0.54}_{-0.43}$ & $<5.9^{ }_{ }$ & $<10^{ }_{ }$ & \nodata & \nodata & \nodata & ACIS \\ 
HHH86-38 & $201.599020$ & $-42.900292$ & $-9.72$ & $1.60^{+0.31}_{-0.27}$ & $<2.5^{ }_{ }$ & $<4.3^{ }_{ }$ & \nodata & \nodata & \nodata & ACIS \\ 
HHH86-33 & $201.317717$ & $-42.848127$ & $-9.68$ & $0.87^{+0.34}_{-0.27}$ & $<3.9^{ }_{ }$ & $<6.7^{ }_{ }$ & \nodata & \nodata & \nodata & ACIS \\ 
Fluffy & $199.545360$ & $-44.157251$ & $-9.67$ & $2.99^{+0.85}_{-0.71}$ & $<5.9^{ }_{ }$ & $<10^{ }_{ }$ & $0.8^{+2.1}_{-0.3}$ & $<13^{ }_{ }$ & $<16.7$ & ACIS-S \\ 
KV19-280 & $201.334631$ & $-42.985827$ & $-9.66$ & $2.47^{+0.50}_{-0.44}$ & $<1.2^{ }_{ }$ & $<2.1^{ }_{ }$ & $1.9^{+1.4}_{-0.8}$ & $<1.1^{ }_{ }$ & $<2.1$ & ACIS \\ 
KV19-273 & $201.304097$ & $-42.989558$ & $-9.59$ & $3.12^{+0.96}_{-0.79}$ & $<0.9^{ }_{ }$ & $<1.5^{ }_{ }$ & $2.0^{+2.3}_{-1.0}$ & $<0.7^{ }_{ }$ & $<1.6$ & ACIS \\ 
HGHH-43 & $201.269935$ & $-43.160808$ & $-9.57$ & $0.91^{+0.32}_{-0.27}$ & $<2.7^{ }_{ }$ & $<4.8^{ }_{ }$ & \nodata & \nodata & \nodata & ACIS \\ 
PFF-GC056 & $201.386638$ & $-42.940098$ & $-9.52$ & $1.53^{+0.42}_{-0.35}$ & $10.1^{+2.1}_{-2.2}$ & $17.6^{+3.7}_{-3.9}$ & \nodata & \nodata & \nodata & ACIS-I \\ 
HGHH-44 & $201.382221$ & $-43.322982$ & $-9.52$ & $2.23^{+0.35}_{-0.34}$ & $<5.2^{ }_{ }$ & $<9.1^{ }_{ }$ & \nodata & \nodata & \nodata & EPIC \\ 
HGHH-G066 & $201.263148$ & $-43.050712$ & $-9.49$ & $1.82^{+0.57}_{-0.47}$ & $<0.9^{ }_{ }$ & $<1.6^{ }_{ }$ & \nodata & \nodata & \nodata & ACIS \\ 
R223 & $201.386667$ & $-43.117278$ & $-9.49$ & $2.04^{+0.67}_{-0.55}$ & $<1.1^{ }_{ }$ & $<2.0^{ }_{ }$ & \nodata & \nodata & \nodata & ACIS \\ 
HGHH-G219 & $201.322045$ & $-42.979623$ & $-9.42$ & $2.79^{+0.69}_{-0.58}$ & $<1.2^{ }_{ }$ & $<2.1^{ }_{ }$ & $1.8^{+1.5}_{-0.8}$ & $<1.1^{ }_{ }$ & $<2.3$ & ACIS \\ 
PFF-GC028 & $201.254757$ & $-42.947666$ & $-9.40$ & $1.15^{+0.41}_{-0.34}$ & $<3.1^{ }_{ }$ & $<5.3^{ }_{ }$ & \nodata & \nodata & \nodata & ACIS \\ 
KV19-295 & $201.418464$ & $-43.047600$ & $-9.34$ & $1.61^{+0.38}_{-0.32}$ & $3.1^{+0.6}_{-0.6}$ & $5.4^{+1.1}_{-1.0}$ & \nodata & \nodata & \nodata & ACIS-I \\ 
T17-1664 & $201.407702$ & $-42.941120$ & $-9.28$ & $1.75^{+0.57}_{-0.46}$ & $<0.8^{ }_{ }$ & $<1.4^{ }_{ }$ & \nodata & \nodata & \nodata & ACIS \\ 
HGHH-G359 & $201.385034$ & $-42.980587$ & $-9.26$ & $2.90^{+0.57}_{-0.51}$ & $26.4^{+3.0}_{-3.1}$ & $46.2^{+5.3}_{-5.5}$ & $2.2^{+1.0}_{-0.8}$ & $23^{+10}_{-8}$ & 7.9 & ACIS-I \\ 
HGHH-G251 & $201.452204$ & $-42.961431$ & $-9.23$ & $1.33^{+0.44}_{-0.35}$ & $6.0^{+2.7}_{-1.9}$ & $10.5^{+4.7}_{-3.3}$ & \nodata & \nodata & \nodata & ACIS-S \\ 
H12$\_194$ & $201.824886$ & $-42.494562$ & $-9.10$ & $3.02^{+0.90}_{-0.74}$ & $<5.1^{ }_{ }$ & $<9.0^{ }_{ }$ & $1.7^{+1.4}_{-0.8}$ & $<5.1^{ }_{ }$ & $<5.1$ & EPIC \\ 
T17-1444 & $201.299889$ & $-42.953690$ & $-9.10$ & $2.57^{+0.71}_{-0.61}$ & $7.0^{+2.0}_{-2.2}$ & $12.1^{+3.5}_{-3.9}$ & $1.2^{+1.1}_{-0.7}$ & $9^{+8}_{-6}$ & 10 & ACIS-I \\ 
T17-1253 & $201.204815$ & $-43.086707$ & $-8.66$ & $2.77^{+0.89}_{-0.73}$ & $1.6^{+0.7}_{-0.7}$ & $2.8^{+0.1}_{-0.1}$ & $1.1^{+0.9}_{-0.6}$ & $2.4^{+0.9}_{-0.6}$ & 5.5 & ACIS-I \\ 
H12$\_141$ & $202.112521$ & $-43.267474$ & $-9.70$ & $1.25^{+0.22}_{-0.20}$ & \nodata & \nodata & \nodata & \nodata & \nodata & \nodata \\ 
HGHH-41 & $201.162357$ & $-43.335133$ & $-9.66$ & $2.39^{+0.37}_{-0.35}$ & \nodata & \nodata & $1.1^{+1.8}_{-0.4}$ & \nodata & \nodata & \nodata \\ 
\enddata
\tablenotetext{a}{ID for object \citep{2022ApJ...929..147D}.}
\tablenotetext{b}{Absolute $V$-band magnitude \citep{2022ApJ...929..147D}.}
\tablenotetext{c}{Dynamical $V$-band $M/L_{V}$ \citep{2022ApJ...929..147D}.}
\tablenotetext{d}{1-10 keV unabsorbed X-ray flux or upper limit.}
\tablenotetext{e}{1-10 keV X-ray luminosity or upper limit.}
\tablenotetext{f}{Inferred dynamical black hole mass \citep{2022ApJ...929..147D}.}
\tablenotetext{g}{$L_{bol}/L_{edd}$ upper limit (for non-detections) or measurement under the assumption that the X-ray emission represents an active galactic nucleus.}
\tablenotetext{h}{Upper limit or estimate of the gas density $n$ in vicinity of central black hole assuming the mass given and radiatively inefficient Bondi accretion.}
\tablenotetext{i}{Instrument used: XMM-Newton/EPIC, Chandra/ACIS-I, or Chandra/ACIS-S.
``ACIS" denotes stacked Chandra ACIS-I and ACIS-S upper limits.}

\end{deluxetable*}

\tablewidth{0pt}
\begin{deluxetable*}{lllllllll}
\label{table:fits}
\tablecaption{Spectral Fit Properties of Identified Sources}
\centering
\tabletypesize{\footnotesize}
\tablehead{
\colhead{ID}& \multicolumn{4}{c}{Free absorption model} & \multicolumn{4}{c}{Fixed absorption model} \\
\colhead{} & \multicolumn{1}{c}{$N_{\mathrm{H}}$\tablenotemark{a}}  & \colhead{$\Gamma$\tablenotemark{b}} & \colhead{Flux\tablenotemark{c}} & \colhead{Fit Stat./DoF\tablenotemark{d}} & \multicolumn{1}{c}{$N_{\mathrm{H}}$\tablenotemark{a}} & \colhead{$\Gamma$\tablenotemark{b}} & \colhead{Flux\tablenotemark{c}} & \colhead{Fit Stat./DoF\tablenotemark{d}} \\
\colhead{} & \multicolumn{1}{c}{($10^{22}$ cm$^{-2}$)}  & \colhead{}  & \colhead{($10^{-15}$ erg s$^{-1}$ cm$^{-2}$)}  & \colhead{} & \multicolumn{1}{c}{($10^{22}$ cm$^{-2})$} & \colhead{} & \colhead{($10^{-15}$ erg s$^{-1}$ cm$^{-2}$)} & \colhead{} }
\startdata
HGHH92-C21  &  $0.23^{+0.21}_{-0.17}$  &  $1.2^{+0.3}_{-0.3}$  &  $21.5^{+2.9}_{-3.0}$  &  $1.1$$(45.4/42)$ & [0.024] &  $0.9^{+0.1}_{-0.1}$  &  $22.8^{+3.0}_{-3.3}$  &  $1.2 (49.7/43)$    \\    
T17-1511  &  $0.53^{+0.30}_{-0.23}$  &  $2.1^{+0.5}_{-0.4}$  &  $15.9^{+2.4}_{-2.4}$  &  $1.0 (28.9/28)$ & [0.024] &  $1.3^{+0.2}_{-0.2}$  &  $18.0^{+2.7}_{-3.0}$  &  $1.6 (45.6/29)$    \\    
HGHH-G359  &  $0.28^{+0.18}_{-0.15}$  &  $1.2^{+0.3}_{-0.2}$  &  $26.4^{+3.0}_{-3.1}$  &  $1.0 (49.7/50)$ & [0.024] &  $0.8^{+0.1}_{-0.1}$  &  $28.8^{+3.2}_{-3.4}$  &  $1.2 (59.1/51)$    \\    
T17-1444\tablenotemark{e}  &  $<0.82^{ }_{ }$  &  $1.6^{+0.3}_{-0.3}$  &  $7.0^{+2.0}_{-2.2}$  &  $0.8 (106.6/132)$  & [0.024]&  $1.6^{+0.3}_{-0.3}$  &  $6.8^{+2.0}_{-2.2}$  &  $0.8 (106.9/133)$    \\    
HGHH92-C23  &  $0.18^{+0.05}_{-0.05}$  &  $1.7^{+0.1}_{-0.1}$  &  $72.1^{+4.0}_{-4.1}$  &  $1.1 (147.9/138)$ & [0.024] &  $1.4^{+0.1}_{-0.1}$  &  $75.6^{+4.1}_{-4.2}$  &  $1.3 (179.8/139)$    \\    
HGHH-07\tablenotemark{e}  &  $0.31^{+0.30}_{-0.19}$  &  $2.4^{+0.6}_{-0.5}$  &  $8.5^{+1.6}_{-1.6}$  &  $1.0 (129.7/128)$ & [0.024] &  $1.7^{+0.2}_{-0.2}$  &  $8.7^{+1.0}_{-1.1}$  &  $1.1 (137.0/129)$    \\    
K-029  &  $0.18^{+0.09}_{-0.08}$  &  $1.6^{+0.2}_{-0.2}$  &  $44.9^{+3.6}_{-3.7}$  &  $0.9 (72.8/81)$ & [0.024] &  $1.3^{+0.1}_{-0.1}$  &  $47.0^{+3.7}_{-3.9}$  &  $1.0 (84.6/82)$    \\    
K-034  &  $<0.16^{ }_{ }$  &  $1.1^{+0.2}_{-0.2}$  &  $77^{+10}_{-10}$  &  $0.8 (29.5/39)$  & [0.024] &  $1.1^{+0.1}_{-0.1}$  &  $76.9^{+9.1}_{-9.9}$  &  $0.7 (29.5/40)$    \\    
HGHH-12\tablenotemark{e}  &  $<0.46^{ }_{ }$  &  $1.7^{+0.7}_{-0.5}$  &  $8.0^{+2.3}_{-2.2}$  &  $0.9 (97.2/107)$ & [0.024] &  $1.5^{+0.3}_{-0.3}$  &  $8.3^{+1.5}_{-1.6}$  &  $0.9 (97.9/108)$    \\    
T17-1412  &  $0.12^{+0.05}_{-0.04}$  &  $1.2^{+0.1}_{-0.1}$  &  $186^{+9}_{-9}$  &  $0.9 (160.2/179)$ & [0.024] &  $1.1^{+0.1}_{-0.1}$  &  $192^{+9}_{-9}$  &  $1.0 (173.9/180)$    \\    
HHH86-34  &  $0.12^{+0.03}_{-0.03}$  &  $1.6^{+0.1}_{-0.1}$  &  $225^{+16}_{-17}$  &  $1.1 (109.5/104)$ & [0.024] &  $1.3^{+0.1}_{-0.1}$  &  $246^{+16}_{-17}$  &  $1.3 (140.7/105)$    \\    
WHH-22\tablenotemark{e}  &  $<0.19^{ }_{ }$  &  $2.1^{+0.9}_{-0.4}$  &  $1.9^{+0.3}_{-0.4}$  &  $0.9 (103.1/113)$ & [0.024]  &  $2.2^{+0.5}_{-0.4}$  &  $1.9^{+0.3}_{-0.4}$  &  $0.9 (103.3/114)$    \\    
HHH86-37\tablenotemark{e}  &  $0.26^{+0.31}_{-0.19}$  &  $2.2^{+0.7}_{-0.5}$  &  $7.7^{+1.7}_{-1.7}$  &  $0.9 (121.5/136)$ & [0.024] &  $2.2^{+0.7}_{-0.5}$  &  $8.3^{+2.1}_{-2.1}$  &  $0.9 (125.7/137)$    \\    
PFF-GC056  &  $<0.72^{ }_{ }$  &  $1.3^{+0.5}_{-0.4}$  &  $10.1^{+2.1}_{-2.2}$  &  $0.8 (16.5/20)$ & [0.024] &  $1.0^{+0.2}_{-0.2}$  &  $11.4^{+2.1}_{-2.4}$  &  $0.9 (18.7/21)$    \\    
KV19-295\tablenotemark{e}  &  $<0.21^{ }_{ }$  &  $1.4^{+0.6}_{-0.2}$  &  $3.1^{+0.6}_{-0.6}$  &  $0.8 (89.2/106)$ & [0.024] &  $1.4^{+0.3}_{-0.3}$  &  $3.0^{+0.5}_{-0.5}$  &  $0.8 (89.2/107)$    \\    
HGHH-G251  &  $0.57^{+0.44}_{-0.30}$  &  $3.5^{+1.4}_{-0.9}$  &  $6.0^{+2.7}_{-1.9}$  &  $1.8 (28.7/16)$ & [0.024] &  $1.3^{+0.3}_{-0.3}$  &  $5.4^{+0.9}_{-1.1}$  &  $2.4 (40.6/17)$   \\  
\enddata
\tablenotetext{a}{Inferred $N_{\mathrm{H}}$ for free fit, or foreground assumed for fixed fit.}
\tablenotetext{b}{Best-fitting photon index for absorbed power-law fit.}
\tablenotetext{c}{Unabsorbed 1--10 keV X-ray flux.}
\tablenotetext{d}{Fit statistic per DoF.}
\tablenotetext{e}{Source fit with Cash statistics.}
\tablecomments{Spectral properties of detected from Table 1 with sufficient counts for a spectral fit. Both free $N_H$ and fixed (to foreground) $N_H$ fits are given.}
\end{deluxetable*}

\clearpage
\end{document}